\documentclass[aps,prl,reprint,groupedaddress]{revtex4-2}
\usepackage{amsmath}
\usepackage{amsmath}
\usepackage{graphicx}
\usepackage{dcolumn}
\usepackage{bm}

\begin{document}

\preprint{APS/123-QED}

\title{Orbital Optical Chirality as the Origin of Vortex Dichroism}

\author{Yoshito Y. Tanaka}
 \thanks{ytanaka@es.hokudai.ac.jp}
\author{Keiji Sasaki}%
 \homepage{sasaki@es.hokudai.ac.jp}
\affiliation{%
Research Institute for Electronic Science, Hokkaido University, Sapporo 001-0020, Japan
}%

\date{\today}

\begin{abstract}
Optical chirality quantifies the geometrical twisting of electromagnetic fields and underlies chiral light–matter interactions, yet its conventional formulation captures only the spin-associated chiral geometry of light. We introduce orbital optical chirality and derive its continuity equation, revealing physical properties distinct from those of angular momentum. For vortex beams, spin and orbital optical chiralities follow the spin and orbital indices, respectively. Applied to a twisted nanorod dimer, orbital optical chirality gives rise to vortex dichroism through quadrupolar hybridized modes, while dipolar modes exhibit only circular dichroism. These results establish a unified framework for optical chirality and provide access to chiral geometries beyond those resolved by spin optical chirality alone.  

\end{abstract}

\maketitle


Chirality—the three-dimensional geometric property of an object that cannot be superimposed onto its mirror image—is ubiquitous in physical, chemical, and biological systems, underpinning phenomena ranging from the homochirality of biomolecules to broken symmetries in condensed matter and particle physics. Characterizing and manipulating chiral matter using tailored optical fields has therefore become a central topic in molecular spectroscopy and photonic engineering \cite{ref1, ref2, ref3}. In 1964, Lipkin identified a family of conserved quantities in free electromagnetic fields, later known as Lipkin’s zilches \cite{ref4}. In 2010, Tang and Cohen demonstrated that one of these quantities directly governs the asymmetry in the dipolar excitation rate of small chiral molecules, thereby establishing the modern concept of optical chirality (OC) \cite{ref5}. Since then, OC has become the standard quantity for describing circular dichroism (CD) and has been widely employed in enhanced chiroptical spectroscopy and the design of superchiral optical fields \cite{ref6, ref7, ref8}.

Despite its remarkable success, the established OC framework is fundamentally restricted to the spin degree of freedom of light, a limitation explicitly reflected in its continuity equation \cite{ref9, ref10, ref11}. By contrast, the angular momentum of light consists of two independent components: spin angular momentum (spin AM), associated with the local rotation of the electromagnetic field, and orbital angular momentum (orbital AM), associated with the helical phase structure of optical vortex beams \cite{ref12}. Although OC and AM belong to different symmetry classes—the former being parity-odd (P-odd) and time-even (T-even), and the latter parity-even (P-even) and time-odd (T-odd)—the continuity equation for OC contains only spin AM as its flux density. Consequently, the established OC framework captures only one half of the chiral geometries of light: its sign follows the spin index but remains completely insensitive to the orbital index. This gap leaves  chiroptical interactions governed by the orbital degree of freedom of light without a corresponding description in terms of optical chirality.

This limitation has recently been highlighted by experiments on vortex dichroism (VD), also referred to as helical or chiral dichroism, namely CD-like responses induced by left- and right-handed optical vortex beams interacting with chiral matter. VD has been observed in a variety of systems ranging from molecules to artificial microstructures \cite{ref13, ref14, ref15, ref16, ref17}. In particular, our recent experiments on a single twisted gold nanorod dimer—a well-defined model chiral system—revealed a quadrupolar dichroic response whose sign is determined solely by the orbital index rather than the spin index \cite{ref18}. This observation exposes a fundamental limitation of the existing OC framework and directly motivates the present work.

In this Letter, we introduce orbital optical chirality (orbital OC), a new P-odd and T-even pseudoscalar complementary to spin OC. Orbital OC is formulated within the same correlation-based framework as spin OC, but characterizes the geometrical twisting of the electromagnetic-field distribution rather than that of the electromagnetic-field vector. Its continuity equation reveals fundamentally distinct physical properties from those of angular momentum. We further show analytically that orbital OC selectively couples to the quadrupolar hybridized mode of a twisted nanorod dimer, thereby identifying orbital OC as the physical origin of vortex dichroism. These results establish a unified theoretical framework that encompasses spin and orbital OC and provide the missing theoretical basis for chiroptical interactions governed by the orbital degree of freedom of light.

We begin by clarifying the fundamental distinction between angular momentum and optical chirality as physical quantities. Angular momentum—whether spin or orbital—describes the dynamical rotation of the electromagnetic field, whereas optical chirality characterizes its geometrical twisting. This viewpoint is formulated through a unified correlation framework, which reproduces the spin and orbital components of AM and extends naturally to their OC counterparts.

For simplicity, we consider a paraxial optical field propagating along the $z$-axis, with electric field $\bm{E}(r, \theta, z, t)$. Since spin AM characterizes the dynamical rotation of the \textit{electric-field vector}, we consider the correlation between the change in $\bm{E}$ over a time interval $\Delta t$ and the change induced by rotating $\bm{E}$ through an angle $\Delta \theta$,
\begin{equation}
[\bm{E}(t + \Delta t) - \bm{E}(t)] \cdot [\mathrm{R}(\Delta\theta)\bm{E}(t) - \bm{E}(t)],
\end{equation}
where the rotation operator is
\begin{equation}
\mathrm{R}(\theta) = \begin{bmatrix}
\cos\theta & -\sin\theta & 0 \\
\sin\theta & \cos\theta & 0 \\
0 & 0 & 1
\end{bmatrix}.
\end{equation}
Under the conditions $\omega\Delta t \ll 1$ and $\Delta\theta \ll 1$, where $\omega$ is the angular frequency, Eq.~(1) becomes
\begin{equation}
\begin{aligned}
\left.\frac{\partial \bm{E}(t)}{\partial t} \cdot \frac{\partial [\mathrm{R}(\theta)\bm{E}(t)]}{\partial \theta}\right|_{\theta=0} \Delta t \Delta \theta
= -\left[ \frac{\partial \bm{E}}{\partial t} \times \bm{E} \right]_z \Delta t \Delta \theta.
\end{aligned}
\end{equation}
The time average of Eq.~(3) for a monochromatic field is proportional to the spin AM density, $\varepsilon/4\omega \operatorname{Im}(\bm{E}^* \times \bm{E})$, where $\varepsilon$ denotes the permittivity\cite{ref4, ref9, ref11}. Similarly, for orbital AM, which describes the dynamical rotation of the \textit{spatial field distribution}, the analogous correlation, for $\omega\Delta t \ll 1$ and $\Delta\theta \ll 1$, becomes
\begin{equation}
\begin{aligned}
&[\bm{E}(\theta, t + \Delta t) - \bm{E}(\theta, t)] \cdot [\bm{E}(\theta - \Delta \theta, t) - \bm{E}(\theta, t)] \\
&= -\frac{\partial \bm{E}}{\partial t} \cdot \frac{\partial \bm{E}}{\partial \theta} \Delta t \Delta \theta.
\end{aligned}
\end{equation}
The time average of Eq.~(4) corresponds to the orbital AM density, $\varepsilon/4\omega \operatorname{Im}(\bm{E}^* \cdot \partial \bm{E}/\partial \theta)$ \cite{ref12}. Both AMs share the same symmetry class, being P-even and T-odd.

Optical chirality instead characterizes the geometrical twisting of the electromagnetic field. Since spin OC is obtained by replacing the temporal evolution for spin AM with spatial evolution along the propagation direction $z$, the corresponding correlation becomes
\begin{equation}
\begin{aligned}
& [\bm{E}(z + \Delta z) - \bm{E}(z)] \cdot [\mathrm{R}(\Delta\theta)\bm{E}(z) - \bm{E}(z)] \\
&= -\left[ \frac{\partial \bm{E}}{\partial z} \times \bm{E} \right]_z \Delta z \Delta \theta.
\end{aligned}
\end{equation}
for $k\Delta z \ll 1$ and $\Delta \theta \ll 1$, where $k$ is the wavenumber. The last expression corresponds to the spin OC. By extending this construction to three spatial axes and imposing dimensional consistency, electric--magnetic duality, and handedness reversal between temporal and spatial evolution, we obtain the three-dimensional expression for spin OC as
\begin{equation}
C^{\text{spin}} = \frac{\varepsilon}{2}\bm{E} \cdot \bm{\nabla} \times \bm{E} + \frac{1}{2\mu}\bm{B} \cdot \bm{\nabla} \times \bm{B},
\end{equation}
where $\mu$ denotes the permeability. This expression is identical to the optical chirality introduced by Tang and Cohen \cite{ref5}.

Orbital OC follows naturally by analogy, characterizing the geometrical twisting of the spatial field distribution rather than that of electric-field vector. Accordingly, we consider the analogous correlation
\begin{equation}
\begin{aligned}
&[\bm{E}(\theta, z + \Delta z) - \bm{E}(\theta, z)] \cdot [\bm{E}(\theta - \Delta \theta, z) - \bm{E}(\theta, z)] \\
&\quad = -\frac{\partial \bm{E}}{\partial z} \cdot \frac{\partial \bm{E}}{\partial \theta} \Delta z \Delta \theta,
\end{aligned}
\end{equation}
for $k\Delta z \ll 1$ and $\Delta \theta \ll 1$, where the final expression is proportional to orbital OC. Combining the three axial components, imposing dimensional consistency, electric--magnetic duality, and handedness reversal as for spin OC, and transforming the cylindrical derivatives into Cartesian form, we obtain the following pseudoscalar,
\begin{equation}
\begin{aligned}
&C^{\text{orbit}} = \sum_{i=x,y,z} C_i^{\text{orbit}} \\
&= \sum_{i=x,y,z} \left[ \frac{\varepsilon}{2}\nabla_i \bm{E} \cdot (\bm{r} \times \bm{\nabla})_i \bm{E} + \frac{1}{2\mu}\nabla_i \bm{B} \cdot (\bm{r} \times \bm{\nabla})_i \bm{B} \right].
\end{aligned}
\end{equation}
where $(\bm{r} \times \bm{\nabla})_i = \partial/\partial\theta_i$ denotes differentiation with respect to the rotation angle around the $i$-th axis. Eq.~(8) is the principal result of this Letter. Importantly, $C^{\text{spin}}$ and $C^{\text{orbit}}$ possess identical physical dimensions and belong to the same symmetry class, being P-odd and T-even, and thus constitute the same class of physical observable, yet represent independent components of optical chirality. Orbital OC originates from the twisting of the spatial field distribution and is independent of the polarization state.

We now derive the continuity equations for spin and orbital OC within the same local-balance framework as Poynting's theorem for electromagnetic energy. We treat matter as a continuous distribution of charge density $\rho(\bm{r},t)$ and current density $\bm{j}(\bm{r},t)$. For spin OC, taking the time derivative of Eq.~(6) and using Maxwell's equations yields
\begin{equation}
\frac{\partial C^{\text{spin}}}{\partial t} + \bm{\nabla} \cdot \bm{F}^{\text{spin}} = -\bm{j} \cdot \bm{\nabla} \times \bm{E},
\end{equation}
where
\begin{equation}
\bm{F}^{\text{spin}} = \frac{\varepsilon}{2}\bm{E} \times \frac{\partial \bm{E}}{\partial t} + \frac{1}{2\mu}\bm{B} \times \frac{\partial \bm{B}}{\partial t}.
\end{equation}
Equations (9) and (10), originally derived by Lipkin, describe the conventional OC \cite{ref4,ref5}. Importantly, the electric-field term of $\bm{F}^{\text{spin}}$ is identical to the correlation in Eq.~(3), and for a monochromatic field, is proportional to the spin AM with a factor $\omega^2$. Analogously, taking the time derivative of Eq.~(8) and using Maxwell's equations yields
\begin{equation}
\begin{aligned}
&\frac{\partial C^{\text{orbit}}}{\partial t} + \bm{\nabla} \cdot \bm{F}^{\text{orbit}}  \\
& =\sum_{i=x,y,z} \left[ \frac{1}{2}\bm{j} \cdot [\nabla_i (\bm{r} \times \bm{\nabla})_i]\bm{E} + \frac{1}{2}\bm{E} \cdot [\nabla_i (\bm{r} \times \bm{\nabla})_i]\bm{j} \right].
\end{aligned}
\end{equation}
where
\begin{equation}
\begin{aligned}
\bm{F}^{\text{orbit}} &= -\left[ \frac{\varepsilon}{2}\frac{\partial \bm{E}}{\partial t} \cdot (\bm{r} \times \bm{\nabla})\bm{E} + \frac{1}{2\mu}\frac{\partial \bm{B}}{\partial t} \cdot (\bm{r} \times \bm{\nabla})\bm{B} \right] \\
&\quad - \left[ \frac{\varepsilon}{2}\bm{E} \cdot (\bm{r} \times \bm{\nabla})\frac{\partial \bm{E}}{\partial t} + \frac{1}{2\mu}\bm{B} \cdot (\bm{r} \times \bm{\nabla})\frac{\partial \bm{B}}{\partial t} \right].
\end{aligned}
\end{equation}
The electric-field term of $\bm{F}^{\text{orbit}}$ corresponds to the correlation in Eq.~(4), and, for a monochromatic field, is proportional to the orbital AM. Thus, orbital OC is the genuine orbital counterpart of spin OC, established not merely by mathematical analogy but through an explicit continuity equation.

A remarkable feature immediately follows from Eqs.~(11) and (12). The three contributions on the right-hand side of Eq.~(11) cancel upon summation because $\sum_i [\nabla_i (\bm{r} \times \bm{\nabla})_i] = 0$. Moreover, since the two terms in $\bm{F}^{\text{orbit}}$ correspond to the orbital AM but appear with opposite signs, their sum also vanishes identically. The resulting continuity equation is therefore identically balanced, reflecting the geometrical cancellation between opposite orbital-OC contributions rather than a conventional conservation law. As shown in Eq.~(8), each component $C_i^{\text{orbit}}$ ($i = x, y, z$) can be positive or negative values corresponding to left- and right-handed twisting, while their sum is mathematically zero. This identity has a clear geometrical interpretation. Rather than indicating the absence of orbital OC, it reflects the geometrical constraint among its Cartesian components: twisting around one axis necessarily accompanies opposite-handed twisting around at least one other axis. Accordingly, when one component is enhanced or suppressed through interaction with chiral matter, an opposite-handed component around another axis changes correspondingly to maintain zero total orbital OC. The vanishing of total orbital OC is therefore not a limitation: its individual components can be nonzero and characterize axis-dependent spatial twisting of the field distribution. Such component-selective coupling can give rise to vortex dichroism in chiral light--matter interactions, as demonstrated below.

We evaluate spin and orbital OC for an optical vortex. Near the beam waist, the complex electromagnetic fields of a paraxial Laguerre--Gaussian (LG) beam with radial index $p = 0$ are written as
\begin{equation}
\tilde{\bm{E}} = E_0 \frac{\bm{e}_x + i\sigma \bm{e}_y}{\sqrt{2}} \left( \frac{\sqrt{2}r}{w_0} \right)^{|l|} \exp\left(-\frac{r^2}{w_0^2}\right) e^{il\theta} e^{ikz},
\end{equation}
\begin{equation}
\tilde{\bm{B}} = B_0 \frac{-i\sigma \bm{e}_x + \bm{e}_y}{\sqrt{2}} \left( \frac{\sqrt{2}r}{w_0} \right)^{|l|} \exp\left(-\frac{r^2}{w_0^2}\right) e^{il\theta} e^{ikz},
\end{equation}
where {$\bm{e}_x$ and $\bm{e}_y$ are unit vectors along the $x$ and $y$ axes, respectively}, and $\sigma$ and $l$ denote the spin and orbital indices, respectively. Substituting Eqs.~(13) and (14) into Eq.~(6) and taking the time average, we obtain
\begin{equation}
\langle C^{\text{spin}} \rangle = \sigma k \left( \frac{\varepsilon}{4} |\tilde{\bm{E}}|^2 + \frac{1}{4\mu} |\tilde{\bm{B}}|^2 \right).
\end{equation}
Similarly, the time-averaged $z$-axis component of Eq.~(8) is
\begin{equation}
\langle C_z^{\text{orbit}} \rangle = l k \left( \frac{\varepsilon}{4} |\tilde{\bm{E}}|^2 + \frac{1}{4\mu} |\tilde{\bm{B}}|^2 \right).
\end{equation}
Thus, spin OC and the $z$-axis component of orbital OC are independently determined by the spin index $\sigma$ and the orbital index $l$, respectively, while exhibiting the same functional form. The transverse components satisfy
\begin{equation}
\langle C_x^{\text{orbit}} \rangle + \langle C_y^{\text{orbit}} \rangle = -l k \left( \frac{\varepsilon}{4} |\tilde{\bm{E}}|^2 + \frac{1}{4\mu} |\tilde{\bm{B}}|^2 \right),
\end{equation}
which, together with Eq.~(16), yields $\langle C^{\text{orbit}} \rangle = 0$. This optical-vortex example therefore provides an explicit demonstration of the general properties of orbital OC derived above.

We next examine the respective roles of spin and orbital OC in the interaction between an optical vortex and a chiral nanostructure. Our recent experiments on a twisted nanorod dimer (TND) under quadrupolar excitation revealed orbital-index-dependent VD \cite{ref18}, which cannot be explained within the conventional spin-OC framework and suggests a mechanism governed by orbital OC. As shown in Fig.~1, the TND geometry is specified by the axial displacement $\Delta z_T$ and the twist angle $\Delta\theta_T$ between the two nanorods, and the sign of their product, $\Delta z_T \Delta\theta_T$, determines the handedness of the TND. Hybridization of the plasmon modes of the two nanorods gives rise to a lower-frequency bonding mode (Fig.~1(a)) and a higher-frequency antibonding mode (Fig.~1(b)), with the constituent plasmon modes oscillating out of phase and in phase, respectively.

We consider the hybridized eigenmode arising from the coupling of the quadrupolar modes supported by the two nanorods. As illustrated in Figs.~1(a) and 1(b), it is represented by four dipoles oscillating with the same complex amplitude,
\begin{equation}
\tilde{\bm{p}}_i = \tilde{\alpha}\tilde{E}_{\text{eff}}\bm{n}_i, \quad i = 1,2,3,4,
\end{equation}
where $\tilde{\alpha}$ is the complex polarizability, $\tilde{E}_{\text{eff}}$ is the effective electric-field amplitude driving the hybridized eigenmode, and $\bm{n}_i$ is a unit vector along the orientation of the $i$-th dipole, with $\bm{n}_1 = -\bm{n}_2$ and $\bm{n}_3 = -\bm{n}_4$ reflecting the quadrupolar symmetry. The effective electric-field amplitude is given by
\begin{equation}
\tilde{E}_{\text{eff}} = \frac{1}{4}\sum_{i}\tilde{\bm{E}}(\bm{r}_i) \cdot \bm{n}_i,
\end{equation}
where $\tilde{\bm{E}}(\bm{r}_i)$ is the local electric field at the $i$-th dipole position, $\bm{r}_{1,2} = \bm{\rho}_{1,2} - (\Delta z_T/2)\bm{e}_z$ and $\bm{r}_{3,4} = \bm{\rho}_{3,4} + (\Delta z_T/2)\bm{e}_z$, with $\bm{\rho}_1 = -\bm{\rho}_2$ and $\bm{\rho}_3 = -\bm{\rho}_4$ denoting the transverse positions. We calculate the time-averaged absorbed power of the hybridized mode under monochromatic excitation as
\begin{equation}
A = \frac{\omega}{2}\sum_{i}\operatorname{Im}[\tilde{\bm{E}}^*(\bm{r}_i) \cdot \tilde{\bm{p}}_i] = 2\omega\operatorname{Im}(\tilde{\alpha})|\tilde{E}_{\text{eff}}|^2.
\end{equation}
Because $\operatorname{Im}(\tilde{\alpha})$ is frequency dependent and plasmon hybridization splits the resonance frequencies of the bonding and antibonding modes, each mode can be selectively excited by tuning the incident-light frequency. Accordingly, the following analysis treats the two hybridized modes independently.

\begin{figure}[t]
\includegraphics[width=\linewidth]{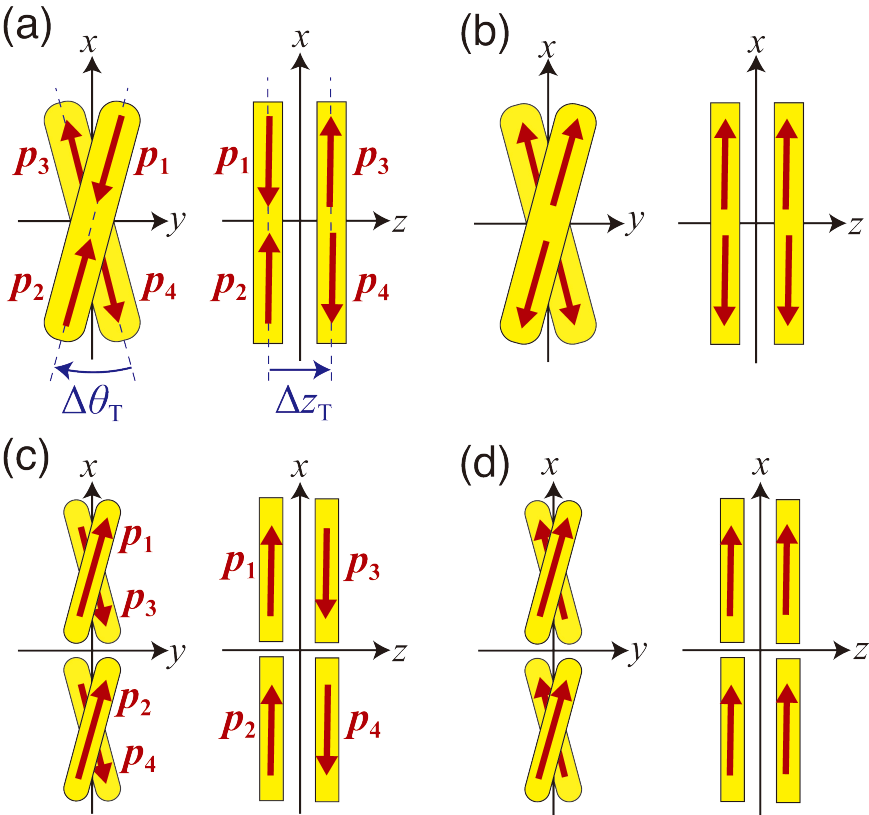}
\caption{\label{fig:epsart} Geometry of twisted nanorod dimers with axial displacement $\Delta z_{\text{T}}$ and twist angle $\Delta\theta_{\text{T}}$. Red arrows represent localized electric dipoles ($\bm{p_1}$--$\bm{p_4}$) forming hybridized eigenmodes arising from the coupling of (a), (b) quadrupolar and (c), (d) dipolar modes: (a), (c) bonding and (b), (d) antibonding modes.   }
\end{figure}

We first evaluate the absorbed power of the bonding mode shown in Fig.~1(a). The TND is assumed to be free to rotate around the $z$ axis, and the absorbed power is averaged over its azimuthal orientation. To facilitate this averaging, we consider an incident field satisfying $\partial |\bm{E}|/\partial \theta = 0$ and $\partial |\bm{E}|/\partial z = 0$, as exemplified by the paraxial LG beam near its waist described by Eq.~(13). To elucidate the role of optical chirality, we consider the limit $k\Delta z_T \ll 1$ and $\Delta\theta_T \ll 1$, and expand Eq.~(20) to second order in $\Delta\theta_T$ and $\Delta z_T$,
\begin{widetext}
\begin{equation}
\begin{aligned}
A_{\text{quad}}^{\text{bond}} &= S_{\text{quad}}\omega\operatorname{Im}(\tilde{\alpha}) \Biggl[ \frac{1}{4}\left( \left| \tilde{E}_x - \frac{\partial \tilde{E}_y}{\partial \theta} \right|^2 + \left| \tilde{E}_y + \frac{\partial \tilde{E}_x}{\partial \theta} \right|^2 \right) \Delta\theta_T^2 + \frac{1}{4}\left( \left| \frac{\partial \tilde{E}_x}{\partial z} \right|^2 + \left| \frac{\partial \tilde{E}_y}{\partial z} \right|^2 \right) \Delta z_T^2 \\
&\qquad + \frac{1}{2}\operatorname{Re}\left[ \tilde{E}_x^* \frac{\partial \tilde{E}_y}{\partial z} - \tilde{E}_y^* \frac{\partial \tilde{E}_x}{\partial z} \right] \Delta z_T \Delta\theta_T - \frac{1}{2}\operatorname{Re}\left[ \frac{\partial \tilde{E}_x^*}{\partial z} \frac{\partial \tilde{E}_x}{\partial \theta} + \frac{\partial \tilde{E}_y^*}{\partial z} \frac{\partial \tilde{E}_y}{\partial \theta} \right] \Delta z_T \Delta\theta_T \Biggr].
\end{aligned}
\end{equation}
\end{widetext}
where $\tilde{E}_x$ and $\tilde{E}_y$ denote the complex $x$- and $y$-components of the incident electric field at an arbitrary dipole. The coefficient $S_{\text{quad}}$, describing the selection rule for quadrupolar excitation, is given by $|\tilde{\bm{E}}(\bm{\rho}) - \tilde{\bm{E}}(-\bm{\rho})|^2 / |2\tilde{\bm{E}}(\bm{\rho})|^2$. For an LG beam, $S_{\text{quad}} = 1$ for odd values of $|l|$ and $S_{\text{quad}} = 0$ for even values, indicating that quadrupolar absorption occurs only for $l = \pm 1, \pm 3, \pm 5, \cdots$. This selection rule arises solely from the rotational symmetry of the quadrupolar excitation and is unrelated to optical chirality. In Eq.~(21), the two terms proportional to $\Delta z_T \Delta\theta_T$ reverse sign with the handedness of the TND and hence represent the chiroptical response. The third term $\operatorname{Re}[\tilde{E}_x^* \partial_z \tilde{E}_y - \tilde{E}_y^* \partial_z \tilde{E}_x]$ and the fourth term $\operatorname{Re}[\partial_z \tilde{E}_x^* \partial_\theta \tilde{E}_x + \partial_z \tilde{E}_y^* \partial_\theta \tilde{E}_y]$ are proportional to the electric-field contributions to the time-averaged spin OC [Eq.~(6)] and the time-averaged $z$-component of orbital OC [Eq.~(8)], respectively. Equation (21) therefore identifies spin OC and orbital OC as the physical origins of the chiroptical absorption. In contrast, neither spin AM nor orbital AM contributes to these handedness-dependent terms. The remaining terms in Eq.~(21), being proportional to $\Delta\theta_T^2$ and $\Delta z_T^2$, are TND handedness independent. For illumination with the LG beam of Eq.~(13), Eq.~(21) becomes
\begin{equation}
\begin{aligned}
A_{\text{quad}}^{\text{bond}} &= S_{\text{quad}}\omega\operatorname{Im}(\tilde{\alpha})|\tilde{\bm{E}}|^2 \Biggl[ \frac{1}{4}(\sigma + l)^2 \Delta\theta_T^2 + \frac{1}{4}k^2 \Delta z_T^2 \\
&\qquad - \frac{1}{2}\sigma k \Delta z_T \Delta\theta_T - \frac{1}{2} l k \Delta z_T \Delta\theta_T \Biggr].
\end{aligned}
\end{equation}
The third and fourth terms originating from spin OC and orbital OC are proportional to the spin index $\sigma$ and orbital index $l$, respectively. These terms change sign with the handedness of geometrical twisting represented by spin OC and orbital OC and therefore give rise to circular dichroism and vortex dichroism, respectively. The first term depends on $(\sigma + l)^2$, the square of the total angular-momentum index, and the second term is index independent; neither exhibits dichroism.

We next analyze the antibonding mode shown in Fig.~1(b), for which the same assumptions and procedure give
\begin{equation}
A_{\text{quad}}^{\text{anti}} = S_{\text{quad}}\omega\operatorname{Im}(\tilde{\alpha}) \left( |\tilde{E}_x|^2 + |\tilde{E}_y|^2 \right) - A_{\text{quad}}^{\text{bond}},
\end{equation}
Equation (23) shows that the bonding and antibonding modes exhibit complementary absorption responses. The first term is the absorption for $\Delta z_T = \Delta\theta_T = 0$, while all terms in $A_{\text{quad}}^{\text{bond}}$, including the spin- and orbital-OC terms, enter with opposite signs. Thus, both circular dichroism and vortex dichroism reverse sign between the bonding and antibonding modes. The spectral separation of these modes makes their opposite orbital-OC responses separately observable.

For comparison, we consider dipolar excitation in each nanorod of the TND. To enable a direct comparison with the quadrupolar case, we employ the same four-dipole model. As illustrated in Figs.~1(c) and 1(d), two TNDs with the same displacement $\Delta z_T$ and twist $\Delta\theta_T$ are arranged side by side, with a dipole excited in each nanorod, thereby forming hybridized eigenmodes within the individual TNDs. The four-dipole configuration is specified by $\bm{n}_1 = \bm{n}_2$, $\bm{n}_3 = \bm{n}_4$, $\bm{r}_{1,2} = \bm{\rho}_{1,2} - (\Delta z_T/2)\bm{e}_z$, and $\bm{r}_{3,4} = \bm{\rho}_{3,4} + (\Delta z_T/2)\bm{e}_z$, with $\bm{\rho}_1 = -\bm{\rho}_2 = \bm{\rho}_3 = -\bm{\rho}_4$. Applying the same assumptions and procedure as in the quadrupolar case, the absorbed power of the bonding modes of the two TNDs shown in Fig.~1(c) is obtained as
\begin{widetext}
\begin{equation}
A_{\text{dip}}^{\text{bond}} = S_{\text{dip}}\omega\operatorname{Im}(\tilde{\alpha}) \left[ \frac{1}{4}\left( |\tilde{E}_x|^2 + |\tilde{E}_y|^2 \right) \Delta\theta_T^2 + \frac{1}{4}\left( \left| \frac{\partial \tilde{E}_x}{\partial z} \right|^2 + \left| \frac{\partial \tilde{E}_y}{\partial z} \right|^2 \right) \Delta z_T^2 + \frac{1}{2}\operatorname{Re}\left[ \tilde{E}_x^* \frac{\partial \tilde{E}_y}{\partial z} - \tilde{E}_y^* \frac{\partial \tilde{E}_x}{\partial z} \right] \Delta z_T \Delta\theta_T \right].
\end{equation}
\end{widetext}
where $S_{\text{dip}}$ denotes the selection-rule factor. When the bonding modes of the two TNDs are coupled to form a collective hybridized mode, $S_{\text{dip}}$ is given by $|\tilde{\bm{E}}(\bm{\rho}) + \tilde{\bm{E}}(-\bm{\rho})|^2 / |2\tilde{\bm{E}}(\bm{\rho})|^2$, where, for an LG beam, $S_{\text{dip}} = 1$ for even values of $|l|$ and $S_{\text{dip}} = 0$ for odd values, indicating that dipolar absorption occurs only for $l = 0, \pm 2, \pm 4, \cdots$. When the bonding modes of the two TNDs are uncoupled, on the other hand, $S_{\text{dip}} = 1/2$ for all $l$, while Eq.~(24) remains unchanged apart from the value of this factor. For illumination with the LG beam of Eq.~(13), Eq.~(24) becomes
\begin{equation}
\begin{aligned}
A_{\text{dip}}^{\text{bond}} = S_{\text{dip}}\omega\operatorname{Im}(\tilde{\alpha})|\tilde{\bm{E}}|^2  \Biggr[ \frac{1}{4}\sigma^2 \Delta\theta_T^2 + \frac{1}{4}k^2 \Delta z_T^2\\ 
\qquad - \frac{1}{2}\sigma k \Delta z_T \Delta\theta_T \Biggr].
\end{aligned}
\end{equation}
For the antibonding dipolar modes shown in Fig.~1(d), the absorbed power is given by
\begin{equation}
A_{\text{dip}}^{\text{anti}} = S_{\text{dip}}\omega\operatorname{Im}(\tilde{\alpha}) \left( |\tilde{E}_x|^2 + |\tilde{E}_y|^2 \right) - A_{\text{dip}}^{\text{bond}}.
\end{equation}
A key difference from the quadrupolar case is that, although the bonding and antibonding modes exhibit the same complementary sign structure, the chiroptical terms proportional to $\Delta z_T \Delta\theta_T$ in Eqs.~(24) and (26) contain only the spin-OC; no orbital-OC appears. For the LG beams, these terms depend only on the spin index $\sigma$, not on the orbital index $l$. Thus, spin OC produces circular dichroism for both dipolar and quadrupolar excitations, whereas orbital OC produces vortex dichroism only for quadrupolar excitation in the present model.

In this Letter, we have established orbital optical chirality as the orbital counterpart of spin optical chirality within a unified correlation-based framework. We derived its continuity equation, uncovered its distinctive physical properties, and identified it as the physical origin of vortex dichroism in twisted nanorod dimers. Together, these results establish optical chirality as a unified framework encompassing both spin and orbital forms.

The relationship between spin and orbital optical chirality is fundamentally different from that between spin and orbital angular momentum. While spin and orbital angular momenta can be mutually converted through light--matter interactions, with only the total angular momentum being conserved \cite{ref19}, no analogous conversion occurs between spin and orbital OC. Spin OC obeys an independent conservation law, whereas total orbital OC remains identically zero as a geometrical constraint. Nevertheless, the individual axial components of orbital OC can undergo compensating changes through light--matter interactions while their sum remains zero. This distinction highlights that optical chirality is not simply another representation of angular momentum but a fundamentally distinct physical quantity.

The distinction between angular momentum and optical chirality becomes particularly transparent for counter-propagating optical vortices. Two counter-propagating LG beams with identical spin and orbital indices have zero net spin AM and zero net orbital AM owing to mutual cancellation, whereas the spin OCs add constructively, as do the orbital OCs, demonstrating that optical chirality remains finite even when spin and orbital AMs vanish. Conversely, when one beam is assigned the opposite spin and orbital indices, the net spin and orbital AM are doubled while both spin and orbital OCs vanish identically. Such beam configurations therefore provide a direct experimental route to verify the independent physical role of optical chirality apart from angular momentum \cite{ref20}.

These fundamental properties are reflected directly in chiral light–matter interactions. The TND analysis further suggests that the appearance of spin and orbital OC is governed by distinct geometrical degrees of freedom of the induced dipole configuration rather than by the multipolar order alone. In the dipolar hybridized modes considered here, only spin OC contributes to the handedness-dependent response, whereas the quadrupolar hybridized modes support both spin and orbital OCs, giving rise to circular and vortex dichroism, respectively. The comparison between the quadrupolar and dipolar excitations suggests an intuitive mechanical interpretation \cite{ref21}: spin OC arises from a spatial rotation of the dipole orientation, whereas orbital OC arises from a rotation of the dipole position about the origin, analogous to intrinsic spin and orbital motion in classical mechanics, respectively. Consequently, spin and orbital OCs provide complementary probes of distinct chiral geometries, with orbital OC providing access to higher-order spatial twisting beyond that resolved by spin OC alone. This principle provides a new design guideline for engineering chiroptical responses through the geometry of light–matter interactions, thereby paving the way for orbital-chirality-selective spectroscopy and the optical control of complex chiral matter.

\begin{acknowledgments}
The authors acknowledge Japan Society for the Promotion of Science (JSPS) KAKENHI (Grant Nos. JP24H00424 and JP22H05132 in Transformative Research Areas (A) “Chiral materials science pioneered by the helicity of light” to Y.Y.T.; JP23H05464, JP25K24583 and JP25K22224 to K.S.), and JSPS International Joint Research Program (JRP-LEAD with UKRI) (Grant No. JPJSJRP20241710); Japan Science and Technology Agency (JST) FOREST Program (Grant No. JPMJFR213O to Y.Y.T.).
\end{acknowledgments}

\bibliography{apssamp}

\end{document}